\documentclass[aps,twocolumn,secnumarabic,amssymb, amsmath,nobibnotes,showpacs]{revtex4}

\usepackage{amssymb,amsmath,amscd}
\usepackage{graphicx,calc,epsfig}
\usepackage{ulem}

\expandafter\ifx\csname package@font\endcsname\relax\else
 \expandafter\expandafter
 \expandafter\usepackage
 \expandafter\expandafter
 \expandafter{\csname package@font\endcsname}%
\fi

\newcommand {\be}{\begin{equation}}
\newcommand {\ee}{\end{equation}}
\newcommand{\ba}{\begin{array}{c}}
\newcommand{\ea}{\end{array}}

\begin{document}
\title{Cyclic universe from Loop Quantum Gravity}%

\author{Francesco Cianfrani$^{\dag}$, Jerzy Kowalski-Glikman$^{\dag}$, Giacomo Rosati$^{\dag}$}%
%\email{francesco.cianfrani@ift.uni.wroc.pl}
\affiliation{$^{\dag}$Institute
for Theoretical Physics, University of Wroc\l{}aw, Pl.\ Maksa Borna
9, Pl--50-204 Wroc\l{}aw, Poland.}
\date{\today}%

\begin{abstract}
\noindent
We discuss how a cyclic model for the flat universe can be constructively derived from Loop Quantum Gravity. This model has a lower bounce, at small values of the scale factor, which shares many similarities with that of Loop Quantum Cosmology. We find that quantum gravity corrections can be also relevant at energy densities much smaller than the Planckian one and that they can induce an upper bounce at large values of the scale factor.  
\end{abstract}


\maketitle%%

The great expectation we have of a quantum theory of gravity is that it will be able to tame the singularities arising in General Relativity. In particular, it should provide us with a consistent, non-singular scenario for the initial stages of our universe. It is hoped that this goal can be achieved in the framework of a quantum cosmological model, constructed by assuming homogeneity and isotropy in the full Quantum Gravity theory, and thus reducing it effectively to a finite dimensional model. Unfortunately, there is no unique way to perform such reduction to quantum cosmology, and the final outcome depends on which path is chosen.

Loop Quantum Gravity (LQG) \cite{Rovelli:2004tv,Thiemann:2007zz,Ashtekar:2004eh} is a framework for a non-perturbative canonical quantization of GR. Loop Quantum Cosmology (LQC) \cite{Bojowald:2011zzb,Ashtekar:2011ni} attempts to quantize the cosmological model by applying the techniques of LQG to {\bf [a]} low-dimensional system obtained by symmetry-reduction of the phase-space of classical GR (minisuperspace). Since the dynamics in minisuperspace is much simpler than in the full theory, this procedure makes it possible to solve some issues of LQG and to realize a consistent quantum scenario for FRW models and for its most relevant anisotropic extensions (Bianchi type I \cite{MartinBenito:2008wx,Ashtekar:2009vc}, II \cite{Ashtekar:2009um} and IX \cite{WilsonEwing:2010rh}). The most important result of these investigations is that the initial singularity is replaced by a {\it bounce} occurring when the energy density reaches the critical value \cite{Singh:2006im,Ashtekar:2006rx}.

A different path to quantum cosmology is proposed in Quantum Reduced Loop Gravity (QRLG) \cite{Alesci:2012md,Alesci:2013xd,Alesci:2013xya}, in which one infers the cosmological sector of LQG directly on a quantum level. This procedure consists of several steps. First, one imposes quantum gauge-fixing conditions by applying the Gupta-Bleuler method, reducing the number of variables and simplifying the expression of geometric operators. Then, one analyzes the quantum Hamiltonian neglecting the non-local terms which arise because of the gauge fixings. It is at this stage that one imposes the homogeneity requirement, since the non-local contributions to the Hamiltonian can be avoided only for homogeneous configurations (e.g. Bianchi I model). Finally, the expectation value of the quantum Hamiltonian on semiclassical states is computed~\cite{Alesci:2014uha,Alesci:2014rra} and it is taken as the generator of the effective semiclassical dynamics.

The models of quantum universe arising from LQC and QRLG differ from the classical one by the presence of two kind of corrections, holonomy and inverse-volume ones, the former being responsible for {\it the bounce replacing the initial singularity}. The magnitude of the corrections is determined by a regulator. In LQC the regulator is treated as a free function of phase-space variables, usually fixed by the demand that the bounce occurs at Planckian energy density~\cite{Ashtekar:2006wn} (the so-called ``improved regularization scheme''). This choice is motivated by the requirement that {\it quantum gravity effects cannot be relevant at length scales much larger than the Planckian ones}. In QRLG, the regulator equals the third root of the inverse number of nodes of the fundamental graph at which the states are based. The most natural choice is to take the regulator to be a constant, since the semiclassical analysis is performed with a non-graph-changing Hamiltonian, which does not add or remove nodes. However, in principle one could look for different definitions of the Hamiltonian and the semiclassical states, which could lead to non-constant regulators and, as a result, to different regularizations.

Within the LQC research program, the choice of a constant regulator, also known as the $\mu_0$ regularization scheme, was adopted in the original proposal (see for instance \cite{Ashtekar:2006uz}), but it has been abandoned, in favor of the improved regularization scheme, due to possible unusual phenomenological implications \cite{Noui:2004gy,Banerjee:2005ga}. However a systematic analysis of the phenomenological downsides was still missing.
In particular, the main reason for discarding a constant regulator is that it implies the energy density at which the bounce occurs to be proportional to the inverse square of the scale factor, so that {\it the quantum gravity effects can be in principle relevant at energy densities much smaller than Planckian one}. This is due to the fact that in LQG a continuous geometry is described by a discrete graph, having a fixed number of nodes, with $SU(2)$ quantum numbers at each link. The physical length scale is determined by both the spin numbers and the fiducial discrete geometry associated with the graph, whose lattice size equals the third root of the inverse number of nodes. As the universe grows, spin numbers grow, the fiducial lattice size stays fixed but the physical lattice size of the geometry keeps growing.

%In this paper, we want to investigate the implications of {\bf [such]} a constant regulator, also known as the $\mu_0$ regularization scheme \cite{Ashtekar:2006uz}, which has been abandoned for its potentially unusual phenomenological implications \cite{Noui:2004gy,Banerjee:2005ga}. In fact, this choice of the regulator implies that the energy density at which the bounce occurs is proportional to the inverse square of the scale factor, so that {\it the quantum gravity effects can be in principle relevant at energy densities much smaller than Planckian one}. This is due to the fact that in LQG a continuous geometry is described by a discrete graph, having a fixed number of nodes, with $SU(2)$ quantum numbers at each link. The physical length scale is determined by both the spin numbers and the fiducial discrete geometry associated with the graph, whose lattice size equals the third root of the inverse number of nodes. As the universe grows, spin numbers grow, the fiducial lattice size stays fixed but the physical lattice size of the geometry keeps growing.

In this work, we are going to investigate the implications of such a decreasing critical energy density in the presence of matter, for which the phenomenological description is given in terms of the ordinary equation of state. 
We thus present for the first time a systematic characterization of a phenomenological model based on the assumption of a constant regulator.
In particular, we will outline how a viable cosmological model can be realized in this framework. Such a model naturally predicts a {\it cyclic universe} oscillating between a minimum scale factor value, taming the initial singularity as in LQC, and a turning point in the future, determined by the cosmological constant (or dark energy) contribution.

Starting from the effective Hamiltonian of gravity coupled to a scalar field presented in~\cite{Alesci:2015nja,Bilski:2015dra}, one can derive, in the isotropic limit, the modified Friedmann and continuity equations. The details of the derivation will be reported in a forthcoming paper~\cite{modcosmLong}, here we present the most important steps.
The semiclassical Hamiltonian is a sum of gravity and scalar field Hamiltonians, $H=H_{gr}+H_{\phi}$, with
\begin{gather}
H_{gr}=-\frac{3}{8\pi G\gamma^{2}}f_{\frac{1}{2}}\sqrt{p}\left(\frac{\sin\left(\mu c\right)}{\mu}\right)^{2}, \\
H_{\phi}=\int d^{3}x\left(\frac{{\cal V}_{0}}{2p^{\frac{3}{2}}}f_{\frac{1}{4}}^{6}\Pi^{2}+\frac{1}{2V_{0}}p^{\frac{3}{2}}V\left(\phi\right)\right).
\end{gather}
In terms of the scale factor, $p \!=\! \ell_{0}^{2}a^{2}$; $\mu$
is equal to $\ell_{pl}/4\pi\gamma\ell_{0}$, $\ell_{pl}$ and $\ell_{0}$
being respectively the Planck length and the fiducial length of the
portion of the universe, whereas ${\cal V}_{0}=\ell_{0}^{3}$ is its fiducial volume;
 $G$ is the Newton's constant, and $\gamma$ is the Immirzi parameter.
It is crucial for the following analysis that $\mu$ does not depend on $p$.

The functions $f_{n}$
are quantum corrections coming from the inverse-volume regularization,
whose expression in terms of $a$ is
\begin{equation}
f_{n}=\frac{1}{2n}a^{2\left(1-n\right)}\left(\left(a^{2}+1\right)^{n}-\left|a^{2}-1\right|^{n}\right).\label{f(a)}
\end{equation}
The phase space variables satisfy the Poisson brackets
\begin{equation}
\left\{ c,p\right\} =\frac{8\pi G\gamma}{3},
~~~~
\left\{\! \phi\left(\mathbf{x}',t\right),\Pi\left(\mathbf{x},t\right) \!\right\} = \delta \!\left(\mathbf{x}' \!\!-\! \mathbf{x}\right).
\end{equation}
The evolution equations are provided by Poisson brackets
of the Hamiltonian constraint: $df/dt=\dot{f}=\left\{ f,H\right\} $.
The relation between the field momentum and its velocity has the form
\begin{equation}
\dot{\phi}=\frac{\lambda V_{0}}{p^{\frac{3}{2}}}f_{\frac{1}{4}}^{6}\Pi.
\end{equation}
Using this one can infer the relation between the field variables and
energy density and pressure comparing the field Hamiltonian
and Lagrangian with those of a perfect fluid $H=\int d^{3}xa^{3}\rho$, $L=\int d^{3}xa^{3}{\cal P}$,
so that
\begin{equation}
\begin{gathered}
\rho=\frac{1}{2\lambda}\left(f_{\frac{1}{4}}^{-6}\dot{\phi}^{2}+V\left(\phi\right)\right),\\
{\cal P}=\frac{1}{2\lambda}\left(f_{\frac{1}{4}}^{-6}\dot{\phi}^{2}-V\left(\phi\right)\right).
\end{gathered}
\label{rho(phi)}
\end{equation}
Using the evolution equation for $p$ and imposing the Hamiltonian constraint $H=0$,
one gets the modified Friedmann equation for the flat ($k=0$) cosmological model (the curvature term can be
added as a phenomenological matter with $w=-1/3$, see the discussion below)
\begin{equation}
\left(\frac{\dot{a}}{a}\right)^{2}=\frac{8\pi G}{3}\rho\left(f_{\frac{1}{2}}-\frac{a^{2}\rho}{6\pi\rho_{pl}}\right)
\label{Friedmann}
\end{equation}
where $\rho_{pl}=m_{pl}/\ell_{pl}\sim 10^{104}\, \mbox{g/cm}^3$ is the Planck density.
By evaluating $\ddot{\phi}=\left\{ \dot{\phi},H\right\} $ one gets
the modified scalar field equation
\begin{equation}
\ddot{\phi}+3\dot{\phi}\frac{\dot{a}}{a}\left(1-2a\frac{d}{da}\ln f_{\frac{1}{4}}\right)+\frac{1}{2}f_{\frac{1}{4}}^{6}\frac{\partial V\left(\phi\right)}{\partial\phi}=0.
\end{equation}
The last equation, re-expressed in terms of the density and pressure
through Eq.(\ref{rho(phi)}) (noting that $\ddot{\phi}=\left(\frac{d}{dt}\dot{\phi}^{2}\right)/2\dot{\phi}$),
gives the modified continuity equation
\begin{equation}
\dot{\rho}=-3\left(\rho+{\cal P}\right)\frac{\dot{a}}{a}\left(1-a\frac{d}{da}\ln f_{\frac{1}{4}}\right).
\label{continuity}
\end{equation}
Using (\ref{Friedmann}) and (\ref{continuity}) one can derive a modified form of the second Friedmann equation.

In what follows we assume that eqs.\ (\ref{Friedmann}) and (\ref{continuity}) can be used to describe evolution of the universe filled with a perfect fluid characterized by energy density $\rho$ and pressure ${\cal P}$.

Notice that, contrary to the standard Friedmann equation, the modified Friedmann equation~(\ref{Friedmann}) is not invariant under rescaling of $a$ by a constant. This is due to the special role that  $a=1$ plays in the QRLG model, being defined as the value of the scale factor at which the physical lattice size is equal to the Planck length.
The inverse-volume corrections governed by the functions $f_n$ are relevant only for small values of the scale factor $a$ and become negligible for $a\gg 1$.

Equations \eqref{Friedmann} and \eqref{continuity} are well-grounded only for $a\gtrsim 1$, the reason being that $a$ equals the spin numbers $j$ of the links of the graph and the results presented in \cite{Alesci:2015nja,Bilski:2015dra} are obtained in the large $j$ limit ($j=O(10)$ at least). Hence, in order to remain within the range of validity of the effective equations \eqref{Friedmann} and \eqref{continuity}, we will require in what follows the lower bounce to occur at $a_-\gtrsim 1$. 

The modified Friedmann equation has a remarkable property that the quantum gravity corrections are strong not only in the case of a Planck size universe, when the scale factor $a$ is small and the matter energy density becomes of order of the Planck one. In fact, the presence of the $a^2$ factor can amplify the otherwise tiny ratio $\rho/\rho_{pl}$ so that the correction term on the right hand side of (\ref{Friedmann}) becomes different from its classical value $1$. This happens if the energy density of matter decreases with $a$ slower than $\rho(a) \sim a^{-2}$, (i.e., for $w<-1/3$.) As explained above this remarkable feature is a direct consequence of the adopted regularization scheme. It is not excluded that the same approach may predict the emergence of the quantum gravity corrections in the large size regions also in different situations, shedding some new light, for example, on the black hole physics and Hawking radiation, making it possible to circumvent the Mathur's no-go theorem \cite{Mathur:2009hf}.

To illustrate the dynamics of the universe described by eqs.\ (\ref{Friedmann}) and (\ref{continuity}) we consider the evolution of the universe filled with phenomenological matter being a mixture of cosmological constant, dust and radiation, with energy density $\rho=\rho_{\Lambda}+\rho_{m}+\rho_{\gamma}$, characterized respectively by the equations of state ${\cal P}_i=w_i \rho_i$, with $w_\Lambda=-1$, $w_m=0$, $w_\gamma=1/3$. Integrating Eq.~(\ref{continuity}) one finds
\begin{equation}
\rho_\Lambda \!\!=\! \rho_0 \Omega_\Lambda, ~~ \rho_m \!\!=\! \rho_0 \Omega_m f^3_{\frac{1}{4}} \! \left(\! \frac{a_0}{a} \!\right)^3 \!\!\!, ~~ \rho_\gamma \!\!=\! \rho_0 \Omega_\gamma f^4_{\frac{1}{4}} \!\left(\!  \frac{a_0}{a} \!\right)^4 \!\!\!,
\label{rho(a)}
\end{equation}
where $\Omega_\Lambda\sim0.7$, $\Omega_m\sim0.3$ and $\Omega_\Lambda\sim10^{-5}$ are the contributions to the cosmological parameters today and $\rho_0 = 3H_{0}^{2}/8\pi G$ is the energy density today, with $H_0\sim10^{-18}s^{-1}$ the current value of the Hubble constant.
$a_0$ is the value of the scale factor today, which must be understood in terms of the value of the scale factor at the quantum regime, $a=1$. In principle we don't know the value of $a_0$. However, in the standard cosmology, for $a<a_{eq}$, $a_{eq}$ being the scale factor at matter-radiation equality, radiation energy density dominates the universe evolution, and $a$ scales as the inverse of the temperature. If we assume that the evolution is essentially classical for $1<a<a_0$, it follows that $a_0 \sim T(a=1)/T_{0}$, $T_0$ being the temperature of radiation today. Then, if we knew the temperature characterizing the quantum regime, at which the physical lattice size is of Planck size, we could estimate $a_0$. Assuming it to be the Planck temperature, $T(a=1)\sim T_{pl}\sim 10^{32} K$, it follows that $a_0\sim10^{32}$.
Using this assumption, from Eq.~(\ref{rho(a)}), we see that at present the quantum correction term in Eq.~(\ref{Friedmann}) is of order $a_0^2t_{pl}^2H_0^2/16\pi^2\sim10^{-62}$ ($G=1/t_{pl}^2\rho_{pl}$, with $t_{pl}\sim 10^{-44} s$), i.e. completely negligible.

Let us take the current state of the universe as a starting point of our investigations and find out what is its future (and past) evolution. Currently, the cosmological constant dominates and it will be even more sizable in the subsequent expansion of the universe. So, to get a qualitative picture of its future evolution, let us assume that both matter and radiation contribution vanish. For $a\gg1$ we get
\begin{equation}
\left(\frac{\dot{a}}{a}\right)^{2}=H_{0}^{2}\,\Omega_{\Lambda}
\left(1-a^{2}\frac{t_{pl}^{2}H_{0}^{2}}{16\pi^{2}}\,\Omega_{\Lambda}
\right)\,,
\end{equation}
which can be solved to give
\begin{equation}
a\left(t\right)=\frac{2ce^{H_{0}\Omega_{\Lambda}^{1/2}t}}{1+c^{2}\Omega_{\Lambda}\frac{t_{pl}^{2}H_{0}^{2}}{16\pi^{2}}e^{2H_{0}\Omega_{\Lambda}^{1/2}t}}\,,
\label{a(t)up}
\end{equation}
with $c$ being an integration constant. We see therefore that the universe undergoes first exponential expansion, then it stops at the transition point, which we call the `upper bounce' and denote $B^{(+)}$, to be followed by a period of exponential contraction which brings the universe essentially back to the today's state. The upper bounce appears at
\begin{equation}\label{5}
a_+\equiv a(t_{B^{(+)}}) = \frac{4\pi}{t_{pl} H_0\Omega_\Lambda}\sim 10^{63}\,.
\end{equation}
To estimate when the upper bounce happens one can solve Eq.~(\ref{a(t)up}) for $a_0=a(t_0)$; it follows
\begin{equation}
e^{H_{0}\Omega_{\Lambda}^{1/2}(t_+-t_0)} \sim \frac{a_+}{a_0} .
\end{equation}

If $a_0\sim10^{32}$ then $t_+-t_0\sim 88/H_0 $, which means that it is going to take about 88 ages of the current universe, i.e., about 2800 billion years before the upper bounce is reached.
In general, for $a_0\sim10^{\alpha}$, $t_+-t_0\sim 2.8(63-\alpha) /H_0$, i.e. the bounce is reached about $2.8(63-\alpha)$ ages of the universe.

After reaching the upper bounce the universe starts contracting again. To see what is the final state of the contraction phase let us assume that the evolution is classical for $a_{eq}<a<a_0$. For $a<a_{eq}$ the evolution is dominated by radiation, so that the total energy density evolves as
\begin{equation}
\rho = \rho_{0} \Omega_\gamma\left( f_{\frac{1}{4}}\frac{a_{0}}{a} \right)^4 \left( 1+ \frac{T_{eq}}{T(a)} \right).
\end{equation}
Substituting this relation into Eq.~(\ref{Friedmann}), we find that the lower bounce occurs at
\begin{equation}
a_- \equiv a(t_{B^{(-)}})= \frac{1}{4\pi} \Omega_\gamma^{1/2} H_0 t_{pl} f_{\frac{1}{2}}^{1/2}(a_-) f_{\frac{1}{4}}^2(a_-) a_0^2.
\end{equation}
Since $f_{\frac{1}{4}}\xrightarrow{a\rightarrow 1} 2^{5/4}$ and $f_{\frac{1}{2}}\xrightarrow{a\rightarrow 1} 2^{1/2}$,
it follows that the bounce occurs at $a_B \gtrsim 1$ if $a_0 \gtrsim 10^{32}$.

\paragraph{Discussion-} We presented a cyclic model of the universe motivated by LQG, in which quantum gravity corrections are responsible not only for the lower bounce taming the initial singularity, but also for the upper bounce, after which the universe enters the contracting phase. This is the main difference between our model and cyclic LQC models, in which the upper bounce can only occur purely classically in the case of the $k=1$ cosmological model \cite{Bojowald:2008ma}.

The present analysis is based on using the expectation value of the Hamiltonian on semiclassical states as the generator of the effective semiclassical dynamics. A more rigorous approach would be to analyze the quantum dynamics of semiclassical states and to infer the behavior of expectation values. Although additional quantum corrections may arise, especially close to the bouncing region, nevertheless we do not expect them to provide significant modifications to the presented scenario. For example, similar investigations in LQC outline how the effective semiclassical dynamics accounts for the relevant quantum corrections \cite{Ashtekar:2006uz}.

The emergence of quantum gravity effects at energy densities much smaller than the Planckian one is the key-point of our investigation. We have seen how they provide significant modifications to the classical behavior (upper bounce) in the presence of matter with $w<-1/3$.
This poses the question whether the standard (slow-roll) inflationary scenario can be accommodated within our framework. Preliminary investigations indicate that it is very unlikely to have a successful inflation. This is due to the fact that to prevent an upper bounce during the inflationary phase, we must make sure that $a^2\rho\ll 1$ at all stages of inflation. Assuming that the number of e-foldings is $N\sim 50$ \cite{Liddle:2000cg}, and the standard mechanism for generation of fluctuations, this results in the condition  $\epsilon\ll 10^{-52}$ for the first slow-roll parameter, which is very unnatural. 

Fortunately, our model shares all the phenomenological features of the standard bouncing cosmologies \cite{Battefeld:2014uga} (see \cite{WilsonEwing:2012pu} for the application to LQC), which provide an alternative to inflation and solve the paradoxes of the standard cosmological model. In our model the bounce occurs smoothly, so that all the issues plaguing bouncing models could be addressed in our framework. In particular, the anisotropic shear does not drive universe evolution if the bounce occurs for $a_-\geq 1$, thus it cannot spoil the presented scenario. 

The behaviour of perturbations is crucial for the viability of the model, both with and without inflation. Their treatment is highly nontrivial in QRLG, since their dynamics is generated by those terms in the Hamiltonian that disappear thanks to homogeneity. Hence, the inclusion of perturbations is the next major step to be made by QRLG. Once addressed, the present model will be ready to be tested against experimental data of precision cosmology.

{\it Acknowledgment}
This work is supported by funds provided by the National Science Center under the agreement
DEC-2011/02/A/ST2/00294, and for JKG also by funds provided by the National Science Center under the agreement 2014/13/B/ST2/04043.

\end{document}